\documentstyle[12pt,psfig]{article}

\begin{document}
\bigskip
\bigskip
\bigskip
\bigskip
\begin{center}\bf\large
    Decay out of a Superdeformed Band 
\end{center}
\bigskip
\begin{center}
     Jian-zhong Gu and H. A. Weidenm\"uller\\
\bigskip
  Max-Planck-Institut f\"ur Kernphysik, Postfach 
103980, D-69029 Heidelberg, Germany

\end{center}
\bigskip
\bigskip
\begin{center}
			      Abstract
\end{center}

Using a statistical model for the normally deformed states and for their 
coupling to a member of the superdeformed band, we calculate the ensemble 
average and the fluctuations of the intensity for decay out of the
superdeformed band and of the intraband decay intensity. We show that 
both intensities depend on two dimensionless variables: The ratio 
$\Gamma^{\downarrow}/\Gamma_S$ and the ratio $\Gamma_N/d$. Here, 
$\Gamma^{\downarrow}$ is the spreading width for the mixing of the 
superdeformed and the normally deformed states, $d$ is the mean level 
spacing of the latter, and $\Gamma_S$ ($\Gamma_N$) is the width for 
gamma decay of the superdeformed state (of the normally deformed states, 
respectively). This parametric dependence differs from the one predicted 
by the approach of Vigezzi et al. where the relevant dimensionless variables 
are $\Gamma_N/\Gamma_S$ and $\Gamma^{\downarrow}/d$. We give analytical 
and numerical results for the decay intensities as functions of the 
dimensionless variables, including an estimate of the error incurred by
performing the ensemble average, and we present fit formulas useful for the 
analysis of experimental data. We compare our results with the approach of 
Vigezzi et al. and establish the conditions under which this approach 
constitutes a valid approximation.

\bigskip


\bigskip
\bigskip
\begin{center}

  e-mail:~~~gu@daniel.mpi-hd.mpg.de
\end{center}
\pagebreak

\section{Introduction}
\label{int}

The study of superdeformed (SD) bands is one of the most active fields of 
nuclear structure studies at high spin \cite{twi86}. The intensities of the 
E2 gamma transitions within a SD band show a remarkable feature: The intraband 
E2 transitions follow the band down with practically constant intensity. At 
some point, the transition intensity starts to drop sharply. This phenomenon 
is referred to as the decay out of a SD band \cite{her87}. It is attributed 
to a mixing of the SD states and the normally deformed (ND) states with equal 
spin. The barrier separating the first and second minima of the deformation 
potential depends on and decreases with decreasing spin $I$. Decay out of the 
SD band sets in at a spin value $I_{0}$ for which penetration through the 
barrier is competitive with the E2 decay within the SD band, see 
Ref.~\cite{abe99}.

The theoretical description of the mixing between SD and ND states uses a 
statistical model for the ND states first proposed by Vigezzi et al. 
\cite{vig90,shi92,shi93}. The ND states to which the SD state is coupled, lie 
several MeV above the ground state. The spectrum of these states is expected 
to be rather complex. It is, therefore, assumed that the ND states can be 
described in terms of random--matrix theory or, more precisely, by the 
Gaussian Orthogonal Ensemble (GOE) of random matrices \cite{vig90}. Likewise, 
the E1 decay of the ND states is calculated within the statistical model. The 
results of this approach have been used to analyze experimental data 
\cite{kru96,kru97,kue97,kho96}. It is one of the aims of such work to 
determine the strength of the coupling between the SD and the ND states and, 
thereby, properties of the barrier separating the first and the second 
minimum of the deformation potential. We mention in passing that recently, 
experimental evidence for non--statistical decay out of the SD band has been 
found \cite{sve99} in $^{60}$Zn, a nucleus very different from the nuclei 
investigated earlier \cite{kru96,kru97,kue97,kho96}.

The model of Ref.~\cite{vig90} was re--examined in Ref.~\cite{wei98}. This 
was done because inspection shows that the formulae derived in 
Ref.~\cite{vig90} apply only in the limit where the electromagnetic decay 
widths $\Gamma_N$ and $\Gamma_S$ for the ND and the SD states are small in 
comparison with the spreading width $\Gamma^{\downarrow}$ which describes the 
mixing of the ND and the SD states. However, analysis of the data 
\cite{kru96,kru97,kue97,kho96} has shown that this condition is not met in 
practice. In the present paper, we follow up the work of Ref.~\cite{wei98}
in which attention was focused on the ensemble average of the intraband
decay amplitude. We evaluate the contribution of the fluctuating part and 
show that it cannot be neglected in situations of practical interest, in 
contrast to the claims made in Ref.~\cite{wei98}. We use the supersymmetry 
approach developed in Ref.~\cite{ver85}. With $d$ the mean level spacing of 
the ND states, we show that the intensities for intraband decay and for 
decay out of the SD band depend on the dimensionless variables 
$\Gamma_N/\Gamma_S$ and $\Gamma^{\downarrow}/d$. We give analytical and 
numerical results for this dependence as well as fit formulas to facilitate 
the analysis of experimental data. We compare our results with those of 
Refs.~\cite{vig90} and \cite{wei98}.

\section{Model}
\label{model}

We denote the first SD state with significant coupling to the ND states
during the E2 decay down the SD band by $|0 \rangle$; its energy by $E_{0}$; 
the ND states having the same spin as the state $|0 \rangle$ by $|j \rangle$ 
with $j=1,\ldots,K$ and $K \gg 1$; their energies by $E_{j}$. The ND states 
decay by statistical E1 emission. We assume that the total E1 decay widths of 
all ND states have the common value $\Gamma_{N}$. The matrix elements $V_{0j}$ 
connecting the SD and the ND states are responsible for decay out of the SD 
band. This situation is illustrated in Fig.1.
We assume that the ND states can be modeled as eigenstates of the GOE. 
Then, the energies $E_j$ follow the GOE distribution, and the $V_{0j}$'s are 
uncorrelated Gaussian distributed random variables with mean value zero and 
common variance $v^{2}$. The spreading width $\Gamma^{\downarrow}$ is defined 
as $\Gamma^{\downarrow}=2\pi\:v^{2}/d$. The limit $K \rightarrow \infty$ is 
taken at the end of the calculation.  
\begin{figure}
\centerline{\psfig{figure=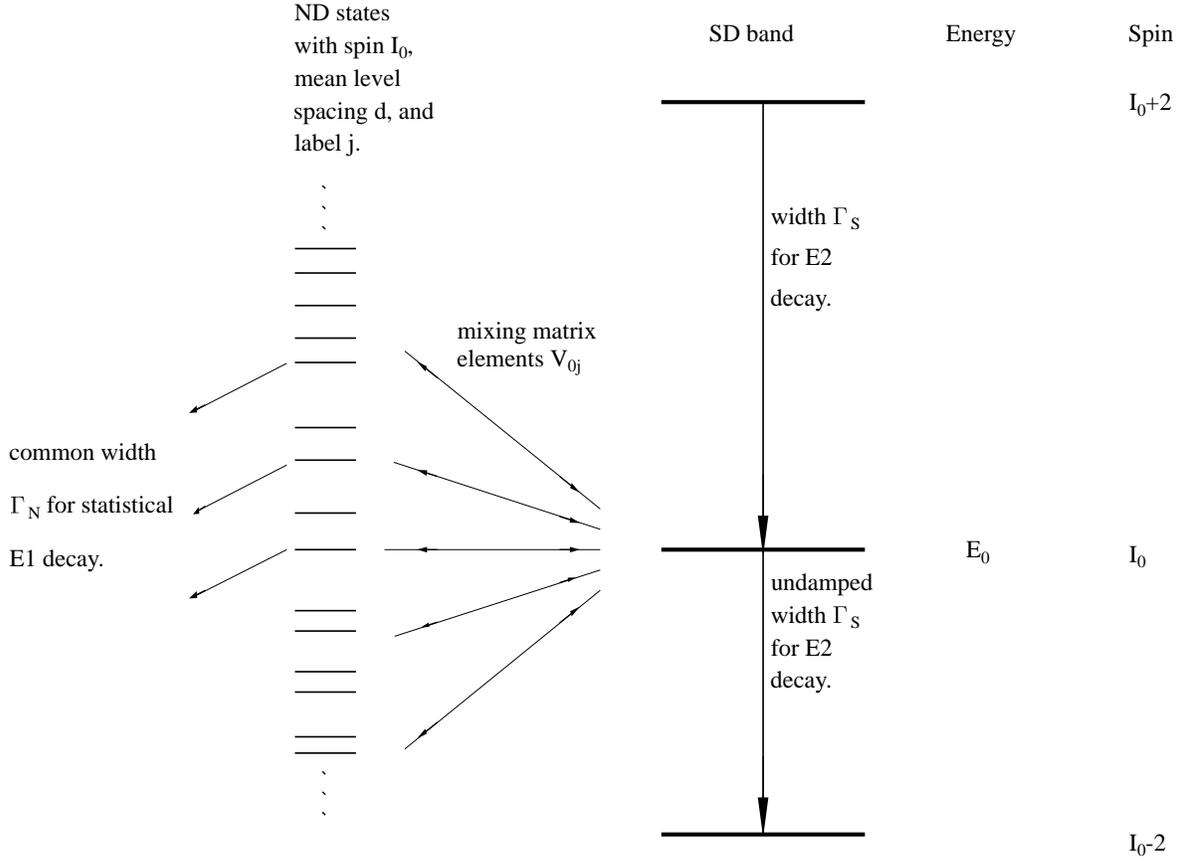,width=19cm,angle=0}}
\caption{ Schematic illustration of the decay out of a SD band due to
admixtures of ND states. The level spacings of ND states are exaggerated in the figure.
}
\label{figmg}
\end{figure}
 
The Hamiltonian $H$ of the system is a matrix of dimension $K+1$ and has the 
form $(j,l = 1, \ldots, K)$
\begin{equation}
\label{ham}
H=
\left(
\begin{array}{cc}
E_{0}&V_{0j}\\
V_{0l}&\delta_{jl}\:E_{j}
\end{array}
\right) \ .
\end{equation}
To H must be added the diagonal width matrix $\Sigma_{SN}$ given by
\begin{equation}
\label{sig}
\Sigma_{SN}=-(i/2)\left(
\begin{array}{cc}
\Gamma_{S}&0\\
0&\delta_{jl}\Gamma_{N}
\end{array}
\right) \ .
\end{equation}
The effective Hamiltonian $H_{\rm eff}$ is given by 
\begin{equation}
\label{eff}
H_{\rm eff} = H + \Sigma_{SN} \ .
\end{equation}
The intraband decay amplitude has the form
\begin{equation}
\label{in}
A_{00}(E)=\gamma_S([E-H_{\rm eff}]^{-1})_{00}\gamma_S \ .
\end{equation}
With $\gamma_{S}^{2}=\Gamma_{S}$, this quantity describes the feeding of the 
SD state from the SD state with the next--higher spin value, and its 
subsequent decay into the SD state with the next--lower spin value. For 
simplicity, we assume that the amplitudes for feeding and decay are both 
given by $\gamma_S$. Similarly, the amplitudes for decay out of the SD band 
are given by
\begin{equation}
\label{ex}
A_{0j}(E)=\gamma_S([E-H_{\rm eff}]^{-1})_{0j}\gamma_N \ ,\: \  j>0 \ ,
\end{equation}
where $\gamma_N^2 = \Gamma_N$. The total intraband decay intensity has 
the form
\begin{equation}
\label{in1}
I_{\rm in}=\frac{1}{2\pi\Gamma_S}\int_{-\infty}^{\infty}|A_{00}(E)|^2dE \ ,
\end{equation}
and the total decay intensity out of the SD band is
\begin{equation}
\label{out1}
I_{\rm out}=\frac{1}{2\pi\Gamma_S}\int_{-\infty}^{\infty}\sum^{K}_{j=1} 
|A_{0j}(E)|^2dE \ .
\end{equation}
The identity 
\begin{equation}
\label{uni}
I_{\rm in} + I_{\rm out} = 1 
\end{equation}
follows from unitarity and completeness. Except for Sections~\ref{pert} and 
\ref{vig}, we will, therefore, focus attention on $I_{\rm in}$.

Both $I_{\rm in}$ and $I_{\rm out}$ vary with the realization of the ensemble 
of random matrices. We are going to calculate the ${\it ensemble \ average}$ 
of both quantities, denoted by a bar. This average involves an average over 
both, the distribution of matrix elements $V_{0j}$ and the distribution of 
eigenvalues $E_j$. In any given nucleus, we deal with ${\it fixed}$ values of 
the $V_{0j}$, and with ${\it fixed}$ positions of the ND states $| j \rangle$. 
In other words, any given nucleus corresponds to a single realization of our 
random--matrix ensemble. The question is: How close to the actual behavior of 
the system will the ensemble average be? To answer this question, we also 
estimate the probability distribution of $I_{\rm in}$. This allows us to 
determine the error incurred by using the ensemble average.

While it is possible to calculate $\overline{I_{\rm in}}$ analytically, the 
calculation of the probability distribution of $I_{\rm in}$ is beyond the 
scope of the supersymmetry technique. Therefore, we use a two--pronged 
approach, employing both the supersymmetry technique and numerical simulation. 
We use analytical results for $\overline{I_{\rm in}}$ as a test for the 
accuracy of our numerical simulation which is then used to estimate the 
probability distribution of $I_{\rm in}$.

\section{Example: Perturbative Approach}
\label{pert}

Various discussions have shown us that application of the GOE to the decay 
out of a superdeformed band involves conceptual difficulties. This fact has 
motivated us to include the present Section in our paper. In this Section, 
we present a simplified version of the GOE approach which can largely be 
dealt with analytically, and which is quite transparent. It is based upon 
a perturbative treatment of the mixing matrix elements $V_{0j}$. From the 
outset, we emphasize that this perturbative treatment is not justified in 
the cases of practical interest, and that our work described in later 
Sections of this paper is not based upon such a perturbative approach. Thus, 
the present Section serves a pedagogical purpose only. We wish to exhibit 
the problems encountered when one applies random--matrix theory to a limited 
data set of a single physical system, and the answers one can give.

We expand the amplitude $A_{0j}$ in powers of $V_{0j}$, keep the first 
non--vanishing term,
\begin{equation}
A_{0j} = \gamma_s (E - E_0 + i \Gamma_S / 2)^{-1} V_{0j} (E - E_j + i
\Gamma_N / 2)^{-1} \gamma_N \ + \ldots \ ,
\label{per}
\end{equation}
and focus attention on the partial width amplitude 
\begin{equation}
\label{pwa}
\gamma_j = V_{0j} (E - E_j + i \Gamma_N / 2)^{-1} \gamma_N \ .
\end{equation}
All random variables of the GOE reside in $\gamma_j$. Moreover, decay out of 
the SD band is (aside from trivial common factors) governed by the quantity 
$\sum_j |\gamma_j|^2$. Therefore, we calculate the first and second moments of 
$\sum_j |\gamma_j|^2$ as GOE averages. Using the statistical independence of 
$V_{0j}$ and $E_j$ and performing first the average over the $V_{0j}$, we find
\begin{equation}
\label{ave1}
\overline{\sum_j |\gamma_j|^2} = v^2 \ \overline{\sum_j \frac{\Gamma_N}
{(E - E_j)^2 + \Gamma_N^2 / 4}} \ .
\end{equation}
To perform the GOE average over the energies $E_j$, we rewrite this expression 
identically as
\begin{equation}
\label{ave2}
\overline{\sum_j |\gamma_j|^2} = v^2 \int {\rm d}E^{\prime} \frac{\Gamma_N}
{(E - E^{\prime})^2 + \Gamma_N^2 / 4} \ \overline{\sum_j \delta(E^{\prime} - 
E_j)} \ .
\end{equation}
By definition, $\overline{\sum_j \delta(E^{\prime} - E_j)} = 1/d$ where $d$ is 
the mean level spacing. Using $\Gamma^{\downarrow} = 2 \pi v^2 / d$, we obtain
\begin{equation}
\label{ave3}
\overline{\sum_j |\gamma_j|^2} = \Gamma^{\downarrow} \ .
\end{equation}
This implies that to lowest order in the $V_{0j}$'s, we have
\begin{equation}
\label{ave4}
\overline{I_{\rm out}} = \Gamma^{\downarrow}/\Gamma_S \ .
\end{equation}
The result~(\ref{ave4}) is remarkable on several counts. First, we observe 
that $\overline{I_{\rm out}}$ depends on the dimensionless variable 
$\Gamma^{\downarrow}/ \Gamma_S$ and not, as Ref.~\cite{vig90} would have us 
expect, on $\Gamma^{\downarrow} /d$ and on $\Gamma_N/\Gamma_S$. Second, we 
find that $\overline{I_{\rm out}}$ is independent of $\Gamma_N$. This fact 
is counter--intuitive because for any given realization of the GOE, the 
quantity $I_{\rm out}$ will surely depend on the magnitude of $\Gamma_N$. 
The independence is caused by our use of first--order perturbation theory 
and by averaging over the eigenvalues $E_j$. This average implies the third 
curious feature of the result~(\ref{ave4}): We smear out the positions of the 
$E_j$'s completely while in the actual experiment, the decay out of a SD band 
will surely depend strongly on where the ND states closest to the SD state 
are located. This expectation can only be reconciled with the 
result~(\ref{ave4}) if the distribution of $I_{\rm out}$ about its mean 
value~(\ref{ave4}) is rather broad. If correct, this statement would imply 
that a determination of $\Gamma^{\downarrow}$ from experimental data using 
the statistical approach would necessarily involve large errors.

To check this point we calculate the variance of $\sum_j |\gamma_j|^2$. A 
straightforward calculation shows that the variance is proportional to 
$(\Gamma^{\downarrow})^2 \ (d/\Gamma_N)$. For the variance of $I_{\rm out}$, 
this implies $\overline{ (I_{\rm out} - \overline{I_{\rm out}} )^2} \sim 
(\Gamma^{\downarrow}/\Gamma_S)^2 \ (d/\Gamma_N)$. This result confirms our 
expectations. For $\Gamma_N \ll d$, i.e., for sharp ND states, the variance of 
$I_{\rm out}$ is bigger than $\bigl (\overline{I_{\rm out}} \bigr )^2$ by the 
huge factor $d/\Gamma_N$. The wider the ND states are, the smaller this 
factor. For $\Gamma_N = d$, the factor is unity and decreases further with 
growing $\Gamma_N/d$. This is physically plausible: For $\Gamma_N \ll d$, 
the decay out of the SD band will sensitively depend on the precise positions 
and decay properties of the ND states, leading to a large uncertainty in the 
prediction afforded by the ensemble average and, hence, in the value of 
$\Gamma^{\downarrow}$ deduced from the data. For obvious reasons this 
dependence of the decay out of the superdeformed band on individual properties 
of the ND states decreases with increasing width of the ND states.

Clearly, lowest--order perturbation theory in $V_{0j}$ is not an adequate way 
of dealing with our problem. Nevertheless, we learn from the example of this 
Section that calculating only the ensemble average of $I_{\rm in}$ or of 
$I_{\rm out}$ is not sufficient. We recall that in the mass region $A \sim 
190$, $d/\Gamma_N$ is typically of the order $10^2$. In this case, we need 
information on the entire probability distribution of $I_{\rm in}$ if we wish 
to assign an error to results obtained from comparing experimental data with 
the GOE average. This point was also emphasized in Ref.~\cite{vig90}. 

\section{Ensemble Average}
\label{ana} 

Following standard procedure in the statistical approach, we write $A_{00}(E)$ 
as the sum of the average part $\overline{A_{00}(E)}$ and the fluctuating part 
$A_{00}^{\rm fluc}(E)$,
\begin{equation}
\label{ana1}
A_{00}(E) = \overline{A_{00}(E)} + A_{00}^{\rm fluc}(E) \ .
\end{equation}
Calculation of the average part is straightforward \cite{wei98} and yields
\begin{equation}
\label{ana2}
\overline{A_{00}(E)} = \gamma_{S}(E - E_0 + i \Gamma_S/2 + 
i \Gamma^{\downarrow}/2)^{-1}\gamma_{S} \ .
\end{equation}
The ensemble average modifies the propagator through the SD state by the 
addition of an imaginary term $i \Gamma^{\downarrow}/2$. This is well known 
from the theory of the optical model in compound--nucleus scattering. The 
decomposition~(\ref{ana1}) entails a corresponding decomposition of 
$\overline{I_{\rm in}}$,
\begin{equation}
\label{ana3}
\overline{I_{\rm in}} =\overline{I_{\rm in}}^{\rm av} +\overline{I_{\rm 
in}}^{\rm fluc} \ ,
\end{equation}
the two terms on the right--hand--side being defined in terms of 
$|\overline{A_{00}(E)}|^2$ and of $|A_{00}^{\rm fluc}(E)|^2$, respectively. 
We have
\cite{wei98} 
\begin{equation}
\label{ana4}
\overline{I_{\rm in}}^{\rm av} = \frac{1}{1 + \frac{\Gamma^{\downarrow}}
{\Gamma_S}} \ .
\end{equation}
We observe that for $\Gamma^{\downarrow} \ll \Gamma_S$, this result agrees 
with the perturbative result for $\overline{I_{\rm out}}$ in 
Section~\ref{pert}.

We turn to the calculation of $\overline{I_{\rm in}}^{\rm fluc}$. In 
Ref.~\cite{wei98}, it was argued that for $\Gamma_N \gg \Gamma_S$ and 
$\Gamma_N \gg \Gamma^{\downarrow}$, this term is negligibly small because 
the ND states decay overwhelmingly by E1 emission. While this assertion is 
certainly qualitatively correct, the question remains: How big is the term 
$\overline{I_{\rm in}}^{\rm fluc}$ in comparison with $\overline{I_{\rm 
in}}^{\rm av}$, and how does it depend on the parameters $\Gamma_S, \Gamma_N, 
\Gamma^{\downarrow}$ and $d$ of our model? To answer these questions, we use 
the supersymmetry formalism \cite{ver85}. We do not reproduce here the 
complete calculation which is lengthy but quite straightforward. It runs 
parallel to that of Ref.~\cite{ver85}. Rather, we describe a shortcut which 
suggests the form of the final result and which also lends plausibility to 
this final result. 

The formalism of Ref.~\cite{ver85} is taylored to compound--nucleus 
scattering. We use the fact that formally, the present problem has much in 
common with compound--nucleus scattering: The ND states are compound--nucleus 
resonances which may decay either by E1 or by E2 gamma emission. The main 
difference to standard compound--nucleus scattering lies in the fact that 
the SD state $|0 \rangle$ acts as a doorway state. Population of the ND 
states from the SD band, and decay of the ND states back into this band, 
both proceed only via intermediate population of the SD state $|0 \rangle$.

To display this similarity, we define the quantity
\begin{equation}
\label{ana5}
S_{00}(E) = 1 - i A_{00}(E) \ .
\end{equation}
We claim that $S_{00}(E)$ can be viewed as a {\it bona fide} $S$--matrix 
element. To make this claim plausible, we consider first the case where 
$V_{0j} = 0$, for all $j$. Then, $S_{00}(E)$ has magnitude one and the form 
\begin{equation}
\label{ana5a}
S_{00}(E) = \frac{E - E_0 - i \Gamma_S / 2} {E - E_0 + i \Gamma_S / 2} \ ,
\ \ {\rm all} \ V_{0j} = 0 \ .
\end{equation}
This is a one--dimensional unitary $S$--matrix describing elastic scattering 
with a resonance located at $E_0$ of width $\Gamma_S$. For $V_{0j} \neq 0$, 
the coupling of the SD state to the ND states and the ability of the latter 
to undergo E1 decay, open additional decay channels. Then, the magnitude of 
$S_{00}$ is smaller than unity, and $S_{00}$ may be viewed as one element of 
a unitary $S$--matrix comprising the E1 decay channels in addition to the E2 
SD band, and displaying the $(K + 1)$ resonances stemming from the SD state
$| 0 \rangle$ and the $K$ ND states $| j \rangle$ with $j = 1, \ldots, K$. 
The actual construction of the other elements of this $S$--matrix is not 
needed, of course, since we are only interested in $S_{00}(E)$. In analogy to 
Eq.~(\ref{ana1}) we write
\begin{equation}
\label{ana6}
S_{00}(E) = \overline{S_{00}(E)} + S_{00}^{\rm fluc}(E) \ .
\end{equation}
From Eq.~(\ref{ana2}) we have 
\begin{equation}
\label{ana7}
\overline{S_{00}(E)} = \frac{E - E_0 - i \Gamma_S/2 + 
i \Gamma^{\downarrow}/2}{E - E_0 + i \Gamma_S/2 + i \Gamma^{\downarrow}/2} \ .
\end{equation}
The fluctuating parts of $A_{00}(E)$ and of $S_{00}(E)$ differ only by the 
factor $(- i)$. Therefore, we can use Eq.~(8.10) of Ref.~\cite{ver85} to 
calculate $\overline{|A_{00}^{\rm fluc}(E)|^2}$ by calculating 
$\overline{|S_{00}^{\rm fluc}(E)|^2}$. The input parameters of 
this equation are specified as follows. The channel indices $a,b,c,d$ are
all equal to 0. The transmission coefficient $T_0$ which couples to the SD 
band is given by
\begin{eqnarray}
&&T_0        = 1 - |\overline{S_{00}(E)}|^2 \nonumber \\
&& \qquad    = \frac{\Gamma_S \Gamma^{\downarrow}}{(E - E_0)^2 + (\Gamma_S 
+ \Gamma^{\downarrow})^2/4} \ .
\label{ana8}
\end{eqnarray} 
This coefficient displays a resonance at $E = E_0$ with width $\Gamma_S 
+ \Gamma^{\downarrow}$. This is due to the fact that the SD state $|0 
\rangle$ is a doorway state for formation of the ND states from and for 
their decay into the SD band. The parameter $\epsilon$ in Eq.~(8.10) of 
Ref.~\cite{ver85} is given by the difference of the energy arguments of two 
scattering amplitudes. The energy arguments of $A_{00}^{\rm fluc}(E)$ and of 
$A_{00}^{\rm fluc}(E)^{*}$ coincide, suggesting that we put $\epsilon = 0$. 
However, since the E1 decay of the ND states is summarily accounted for in 
terms of their common width $\Gamma_N$, an imaginary energy difference 
arises which amounts to the replacement $\epsilon \rightarrow - i \Gamma_N$.
As a result, E1 decay is accounted for by the appearance of an exponential
factor $\exp( - (\pi \Gamma_N / d) (\lambda_1 + \lambda_2 + 2 \lambda))$
under the integral. All transmission coefficients except $T_0$ must be put 
equal to zero. The resulting equation expresses 
$\overline{|A_{00}^{\rm fluc}(E)|^2}$ as a threefold integral over real 
variables $\lambda, \lambda_1, \lambda_2$. For the calculation of 
$\overline{I_{\rm in}}^{\rm fluc}$, we need to integrate in addition over 
energy $E$. This yields eventually
\begin{eqnarray}
&& \overline{I_{\rm in}}^{\rm fluc} = \frac{1}{16\pi\Gamma_S}
\int_{-\infty}^{\infty} {\rm d}E \int_{0}^{\infty} {\rm d}\lambda_{1}
\int_{0}^{\infty} {\rm d}\lambda_{2} \int_{0}^{1} {\rm d}\lambda \nonumber \\
&& \qquad
\times \frac{(1-\lambda) \lambda |\lambda_1 - \lambda_2|} {((1+\lambda_1) 
\lambda_1 (1+\lambda_2) \lambda_2)^{1/2} (\lambda + \lambda_1)^2
(\lambda + \lambda_2)^2} \nonumber \\
&& \qquad
\times \exp [-\frac{\pi \Gamma_N}{d} (\lambda_{1}+\lambda_{2}+2\lambda) ]
\frac{1-T_0 \lambda} {(1+T_0 \lambda_{1})^{1/2} (1+T_{0}\lambda_{2})^{1/2}}
\nonumber \\
&& \qquad 
\times \bigl (\ |\overline{S_{00}(E)}|^2 \ T_0^2 \ \
(\frac{\lambda_1}{1+T_0 \lambda_1} + \frac{\lambda_2}{1+T_0 \lambda_2} +
\frac{2 \lambda}{1-T_0 \lambda})^2 \nonumber \\
&& \qquad 
+ 2 \ T_0^2  \ \ (\frac{\lambda_1 (1+\lambda_1)} {(1+T_0 \lambda_1)^2} + 
\frac{\lambda_2 (1+\lambda_2)}{(1+T_0 \lambda_2)^2} + \frac{2\lambda 
(1-\lambda)}{(1-T_0 \lambda)^2}) \ \bigr ) \ .
\label{ana9}
\end{eqnarray}
The four integrals must be done numerically. Eq.~(\ref{ana9}) coincides with 
the result obtained by straightforward application of the supersymmetry 
approach. We observe that the integrand in Eq.~(\ref{ana9}) is semi-positive
definite. Therefore, $\overline{I_{\rm in}}^{\rm fluc}$ decreases
monotonically with increasing $\Gamma_N/d$, tending to zero as $\Gamma_N/d 
\rightarrow \infty$. This is what we expect on physical grounds \cite{wei98}.
For $\Gamma_N/d = 0$, we have $\overline{I_{\rm in}}^{\rm fluc} = 
(\Gamma^{\downarrow} / \Gamma_S) / ( 1 + \Gamma^{\downarrow} / \Gamma_S )$.
This follows from $I_{\rm in} = 1$ and from Eqs.~(\ref{uni}, \ref{ana3}) 
and (\ref{ana4}).

The appearance of the exponential factor in Eq.~(\ref{ana9}) can be visualized 
in yet another way. Instead of the replacement $\epsilon \rightarrow - 
i \Gamma_N$, we could have put $\epsilon = 0$ but kept in Eq.~(8.10) of 
Ref.~\cite{ver85} the product over transmission coefficients 
\begin{equation}
\label{ana10}
\prod_l \frac{(1 - T_l \lambda)}{(1 + T_l \lambda_1)^{1/2} (1 + T_l 
\lambda_2)^{1/2}} \ .
\end{equation}
We could have argued that this product accounts for both, coupling to the 
SD band via $T_0$, and for E1 decay into a large number of open decay 
channels described by the transmission coefficients $T_l$ with $l \neq 0$. 
Owing to the weakness of the electromagnetic force, we would have $T_l 
\ll 1$ for all $l \neq 0$, although the sum $\sum_{l \neq 0} T_l$ may be 
significant. Excepting $l = 0$, we could then approximate the 
product~(\ref{ana10}) by $\exp( - (1/2) (\sum_l^{\prime} T_l) (\lambda_1 + 
\lambda_2 + 2 \lambda))$. The prime indicates that the term with $l = 0$ 
is omitted. Comparison of this exponential with the one in Eq.~(\ref{ana9}) 
shows that $\sum_l^{\prime} T_l =  2 \pi \Gamma_N / d $. This is a very 
satisfactory result. It is identical to the standard relation connecting 
decay width and sum over transmission coefficients in the theory of 
nuclear reactions, cf. Ref.~\cite{ver85}. Hence, we identify the total 
transmission coefficient $T_N$ for E1 decay as
\begin{equation}
\label{ana11}
T_N = 2 \pi \Gamma_N / d \ .
\end{equation}
We note that in nuclei with mass $A \sim 190$ where $\Gamma_N / d \approx 
10^{-2}$ we have $T_N \approx .05$. This is a rather small value. We will 
see in the next Section that owing to this small value, decay of the ND 
states back into the SD channel is not altogether negligible, in contrast 
to the claim made in Ref.~\cite{wei98}. 

In summary, we have described how to generate an analytical expression for
$\overline{I_{\rm in}}^{\rm fluc}$. As a by--product, we have seen that 
this quantity depends on the input parameters $\Gamma_S, \Gamma_N, 
\Gamma^{\downarrow}$ and $d$ of our model only via $T_0$ and $T_N$, see
Eqs.~(\ref{ana8}) and (\ref{ana11}). Integration over energy $E$ will reduce
the dependence on the dimensionless parameter $T_0$ to a dependence on the 
only surviving dimensionless combination $\Gamma^{\downarrow}/\Gamma_S$ of
parameters contained in $T_0$. Hence, $\overline{I_{\rm in}}^{\rm fluc}$ is
a function of the two dimensionless parameters $\Gamma^{\downarrow}/ 
\Gamma_S$ and $\Gamma_N / d$. This conclusion applies not only to 
$\overline{I_{\rm in}}^{\rm fluc}$ but likewise to all higher moments of
$A_{00}(E)$. Indeed, the supersymmetry approach ~\cite{ver85} yields
formally identical results for these higher moments, except for the fact 
that the dimension of the supermatrices is increased. Thus, we have shown
that the entire probability distribution of $I_{\rm in}$ depends on the 
input parameters of our model only via the two dimensionless variables 
$\Gamma^{\downarrow}/\Gamma_S$ and $\Gamma_N / d$. This is a non--trivial 
result for two reasons. First, from the four input parameters $\Gamma_S, 
\Gamma_N, \Gamma^{\downarrow}$ and $d$ of the model, we can construct three 
independent dimensionless variables. The probability distribution of 
$I_{\rm in}$ depends on only two of them. Second, both dimensionless 
variables are given by the ratio of an electromagnetic and a nuclear quantity. 
Hence, both depend on the ratio of the fine structure constant and the 
strength of the strong interaction.

\section{Results}
\label{res} 

We have calculated the fourfold integral in Eq.~(\ref{ana9}) numerically.
We add the term $\overline{I_{\rm in}}^{\rm av}$, cf. Eq.~(\ref{ana4}), and 
denote the result by $\langle I \rangle$. We have also simulated the 
model of Eqs.~(\ref{ham},\ \ref{sig},\ \ref{in}) numerically. This was done 
by drawing the matrix elements $V_{0j}$ from a Gaussian distribution centered 
at zero with variance $v^2$. The energies $E_j$ were taken from an unfolded 
GOE spectrum with $E_0$ located in the center of the semicircle. Typically, 
we used matrices of dimension $K = 100$ or bigger, and we calculated 
$N = 10^4$ or more realizations. The calculations were simplified by using 
for $A_{00}(E)$ the expression
\begin{equation}
\label{res0}
A_{00}(E) = \frac{\Gamma_S}{E - E_0 +i \Gamma_S / 2 - \sum_{j=1}^K 
\frac{V_{0j} V_{j0}}{E - E_j + i \Gamma_N / 2}} \ .
\end{equation}
We note that in the simulation, we calculate the total intraband intensity 
$I_{\rm in}$ without introducing the distinction between $\overline{I_{\rm 
in}}^{\rm av}$ and $\overline{I_{\rm in}}^{\rm fluc}$. We used $\langle I 
\rangle $ to test the results of the simulation. The width of the probability 
distribution of $I_{\rm in}$ was estimated as follows. With $I(n)$ the value 
of $I_{\rm in}$ obtained in the $n^{\rm th}$ realization ($n = 1, \ldots, N$), 
two sets labelled $s_i$ with $i = 1,2$ were formed depending on whether $I(n) 
< \langle I \rangle$ or $I(n) > \langle I \rangle$, respectively, each set 
containing $N_i$ realizations labelled $\mu_i = 1, \ldots, N_i$. For $i = 
1,2$ we have 
calculated
\begin{equation}
\label{res1}
\sigma_i = \sqrt{ \frac{1}{N_{i}} \sum_{\mu_i}^{N_i} (I(n)- \langle I 
\rangle )^2 } \ .
\end{equation}

The results are shown in Figs.~2(a) to 2(f). Cases (a) to (f) correspond to
six different choices of the ratio $\Gamma_N / d$ as indicated in the Figures. 
In all cases shown, the abscissa gives the ratio $\Gamma^{\downarrow} / 
\Gamma_S$ on a logarithmic scale. The top panels show $\overline{I_{\rm 
in}}^{\rm fluc}$ as calculated from Eq.~(\ref{ana9}). We note that 
$\overline{I_{\rm in}}^{\rm fluc}$ decreases monotonically with increasing 
$\Gamma_N / d$. At the same time, the maximum of $\overline{I_{\rm in}}^{\rm 
fluc}$ shifts towards smaller values of $\Gamma^{\downarrow} / \Gamma_S$. We 
note, moreover, that for $\Gamma_N / d = 10^{-2}$, the peak value is $\approx 
0.6$ and, thus, not small compared to unity. The decay of the ND states back 
into the SD band is negligible only for overlapping ND states, i.e., for 
$\Gamma_N / d \geq 1$. The bottom panels show as solid lines the average 
values $\overline{I_{\rm in}}$ of the total intraband intensity obtained by 
adding $\overline{I_{\rm in}}^{\rm av}$ (Eq.~(\ref{ana4})) and 
$\overline{I_{\rm in}}^{\rm fluc}$ (top panels). All curves decrease 
monotonically with increasing $\Gamma^{\downarrow} / \Gamma_S$. This is 
physically plausible. The value of $\Gamma^{\downarrow} / \Gamma_S$ where 
$\overline{I_{\rm in}}$ equals 1/2 shrinks from $\approx 5 \times 10^4$ for 
$\Gamma_N / d = 10^{-6}$ to $\approx 1 $ for $\Gamma_N / d = 10$. The bottom 
panels also show as error bars the widths of the distributions as estimated 
by the quantities $\sigma_i$ in Eq.~(\ref{res1}). Not surprisingly, the errors 
are biggest when the contribution of $\overline{I_{\rm in}}^{\rm fluc}$ to 
$\overline{I_{\rm in}}$ is largest and shrink with increasing size of 
$\Gamma_N / d$. The error bars reflect directly the role played by individual 
ND states in the E1 decay: For $\Gamma_N / d \gg 1$, the ND states overlap 
strongly, their positions are irrelevant, and the errors are small. The 
opposite situation prevails for $\Gamma_N / d \ll 1$ where the location of the 
ND states closest to the SD state $| 0 \rangle$ is of tantamount importance.

\begin{figure}
\centerline{\psfig{figure=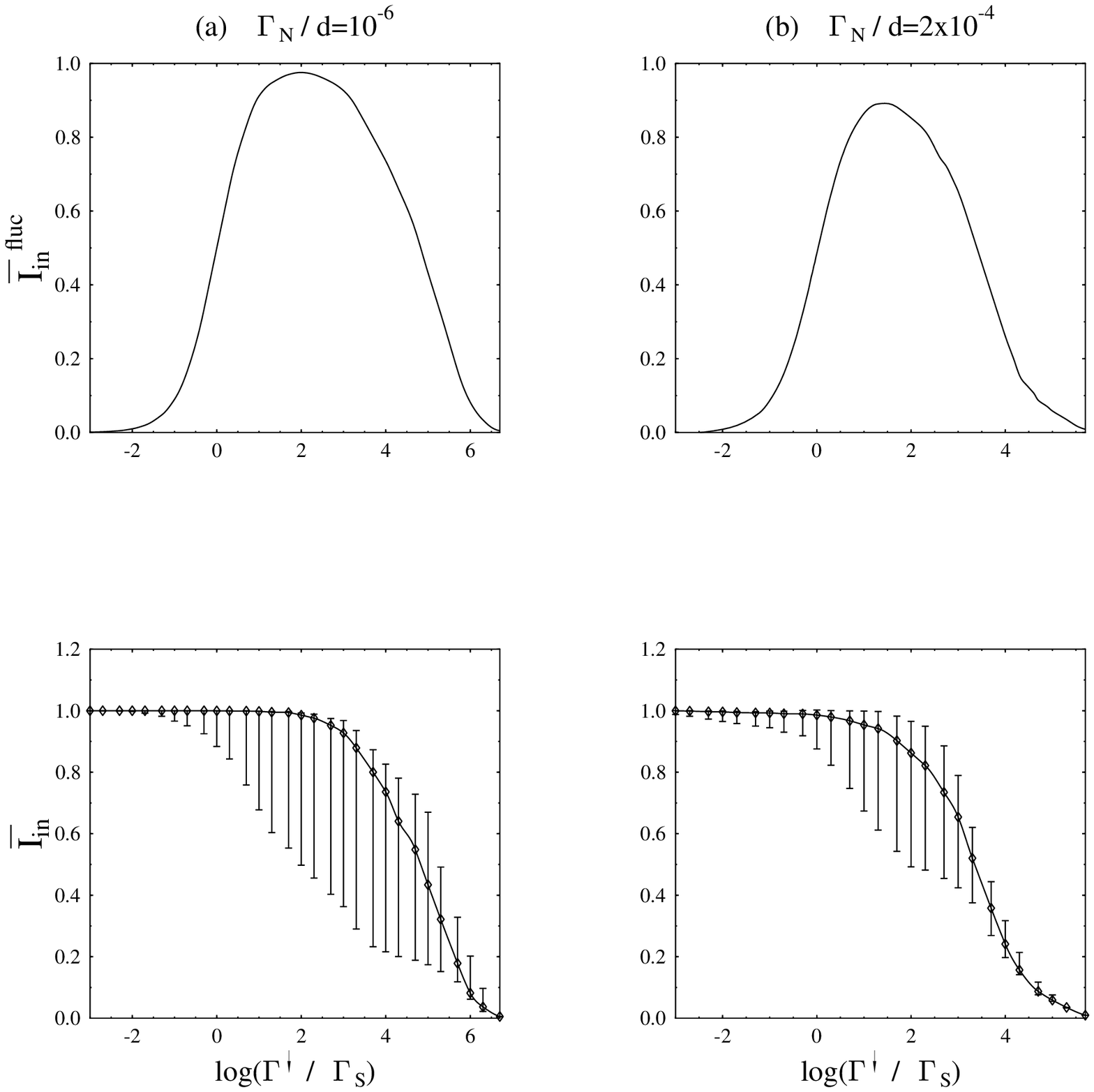,width=19cm,angle=0}}
\end{figure}

\begin{figure}
\centerline{\psfig{figure=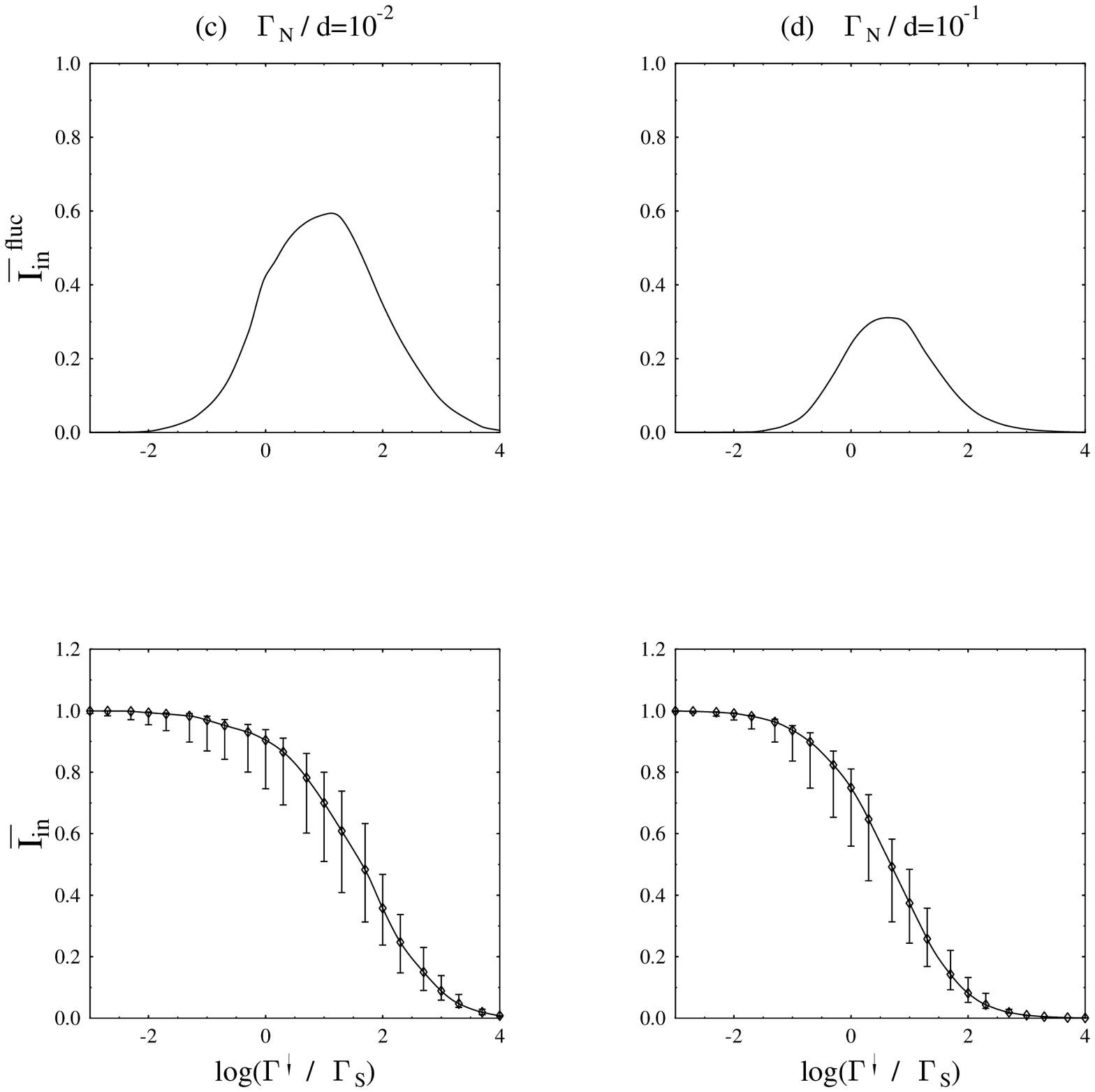,width=17cm,angle=0}}
\end{figure}

\begin{figure}
\centerline{\psfig{figure=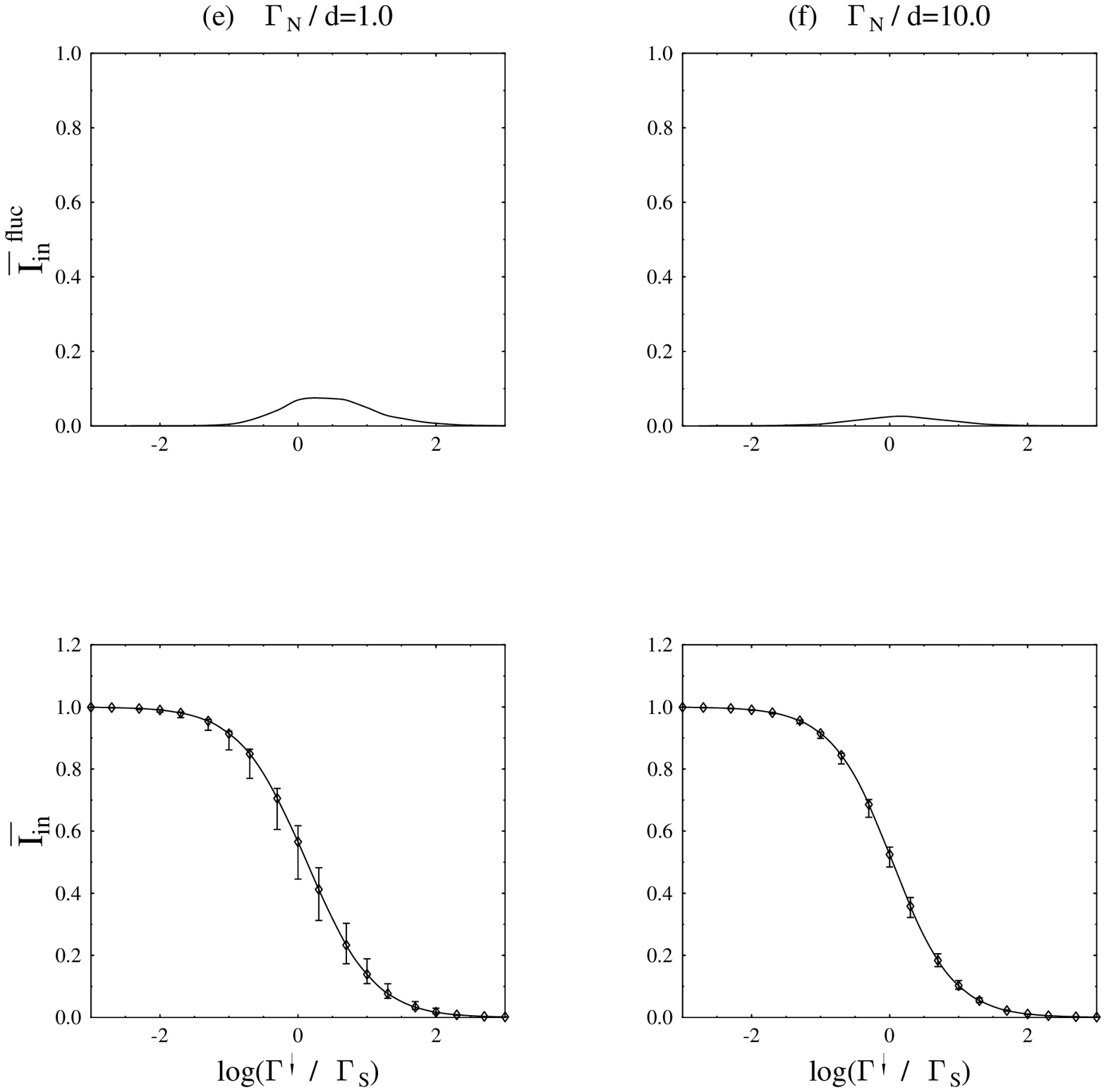,width=16.5cm,angle=0}}
\caption{
The ensemble average $\overline{I_{\rm in}}^{\rm fluc}$ 
(top, abbreviated as $\overline{I}_F$)
and the total average intraband intensity 
$\overline{I_{\rm in}}$ with error bars
due to the statistical uncertainty 
(bottom, abbreviated as $\overline{I}$)
versus logarithm (basis ten) of the ratio $\Gamma^{\downarrow} / 
\Gamma_S$ for several values of $\Gamma_N/d$.}
\label{figmb}
\end{figure}

In order to make our work useful for the analysis of experimental data, we
now present two fit formulas which approximately reproduce the relevant 
behavior of the average intraband decay intensity $\overline{I_{\rm in}}$. 
We fit the curves for $\overline{I_{\rm in}}^{\rm fluc}$ shown in the top 
panels of Figs.~2(a) to 2(f) and find
\begin{equation}
\label{fit}
\overline{I_{\rm in}}^{\rm fluc}
= (1-0.9139(\frac{\Gamma_N}{d})^{0.2172})\times\exp\{-\frac{
(0.4343\ln(\Gamma^{\downarrow}/\Gamma_S)-0.45(\frac{\Gamma_N}{d})^{-0.1303})^2}{
(\Gamma_N/d)^{-0.1477}}\}\ .
\end{equation}
We emphasize that this formula is not based on any theoretical arguments 
and presents the result of an approach based upon trial and error. In 
Fig.~3, we show a comparison between the fit formula~(\ref{fit}) (dotted 
lines) and the calculated values for $\overline{I_{\rm in}}^{\rm fluc}$ 
(solid lines). We observe that the decreasing parts of the curves for 
$\overline{I_{\rm in}}^{\rm fluc}$ (which are particularly relevant for the 
experimental determination of $\Gamma^{\downarrow}$) are particularly well 
reproduced. For fixed values of $\Gamma_N / d$, the difference between the fit 
value and the calculated value is, for any value of $\Gamma^{\downarrow} / 
\Gamma_S$, never bigger than .07 and lies well within the limits of 
uncertainty defined by the error bars in the lower panels of Fig.~2.

\begin{figure}
\centerline{\psfig{figure=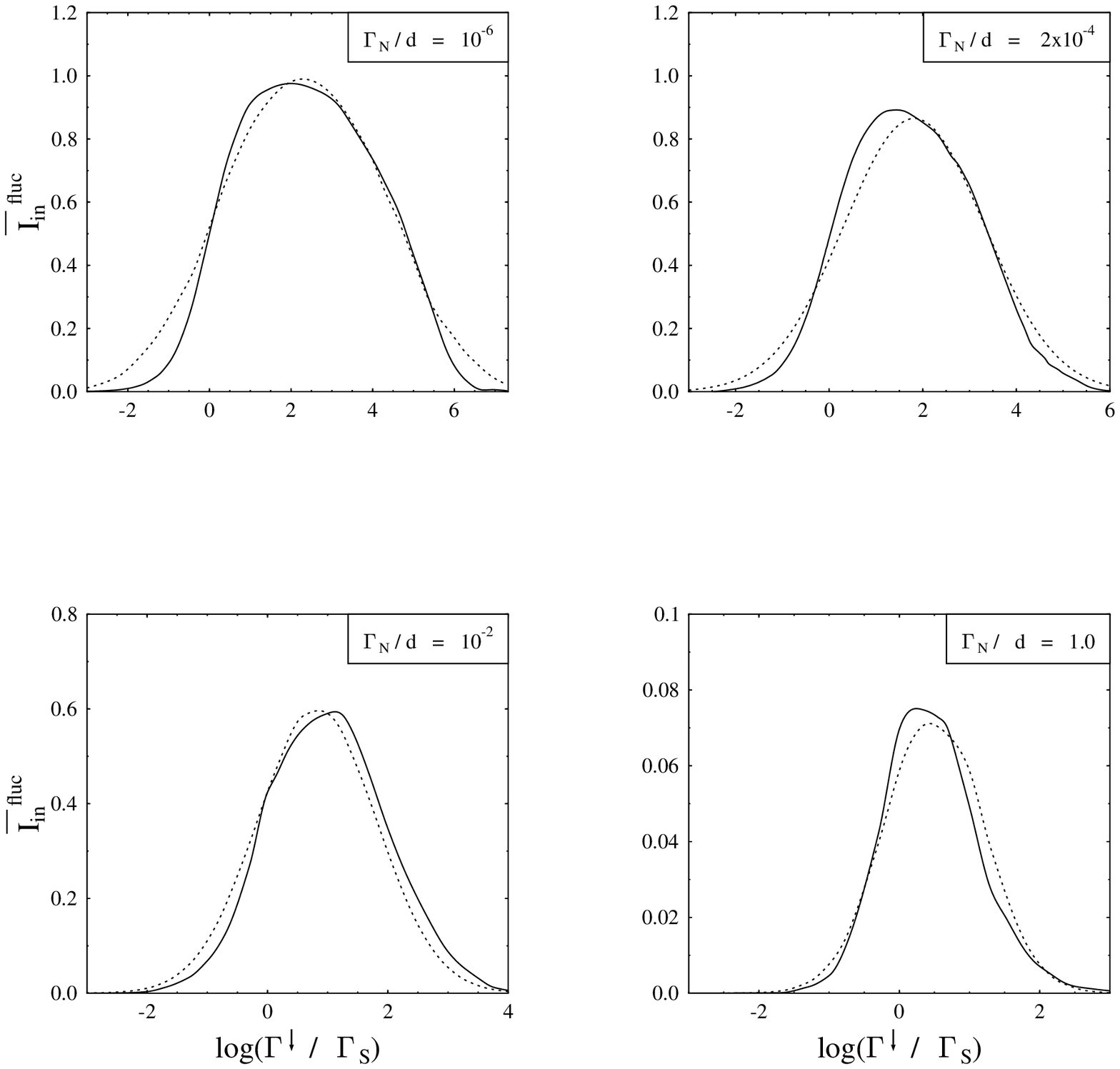,width=17cm,angle=0}}
\caption{
Comparison of the fit formula Eq.~(\ref{fit}) (dotted lines) with the 
ensemble--averaged intraband decay intensity (solid lines)
$\overline{I_{\rm in}}^{\rm fluc}$ 
(abbreviated as $\overline{I}_F$) 
for several values of $\Gamma_N/d $. The basis of logarithm is ten.}
\label{figmf}
\end{figure}

For the lower panels of Figs.~2(a) to 2(f) we fit the values of 
$\Gamma^{\downarrow} / \Gamma_S$ for which, at a given value of $\Gamma_N 
/ d$, the average intraband decay intensity $\overline{I_{\rm in}}$ assumes 
the value $1/2$. We obtain
\begin{equation}
\label{fit1}
\Gamma^{\downarrow}/\Gamma_S=1+0.5429(\Gamma_N/d)^{-0.8947}.
\end{equation}
Formula~(\ref{fit1}) is in good agreement with the calculated values of 
$\Gamma^{\downarrow}/\Gamma_S$, see Fig.~4.

\begin{figure}
\centerline{\psfig{figure=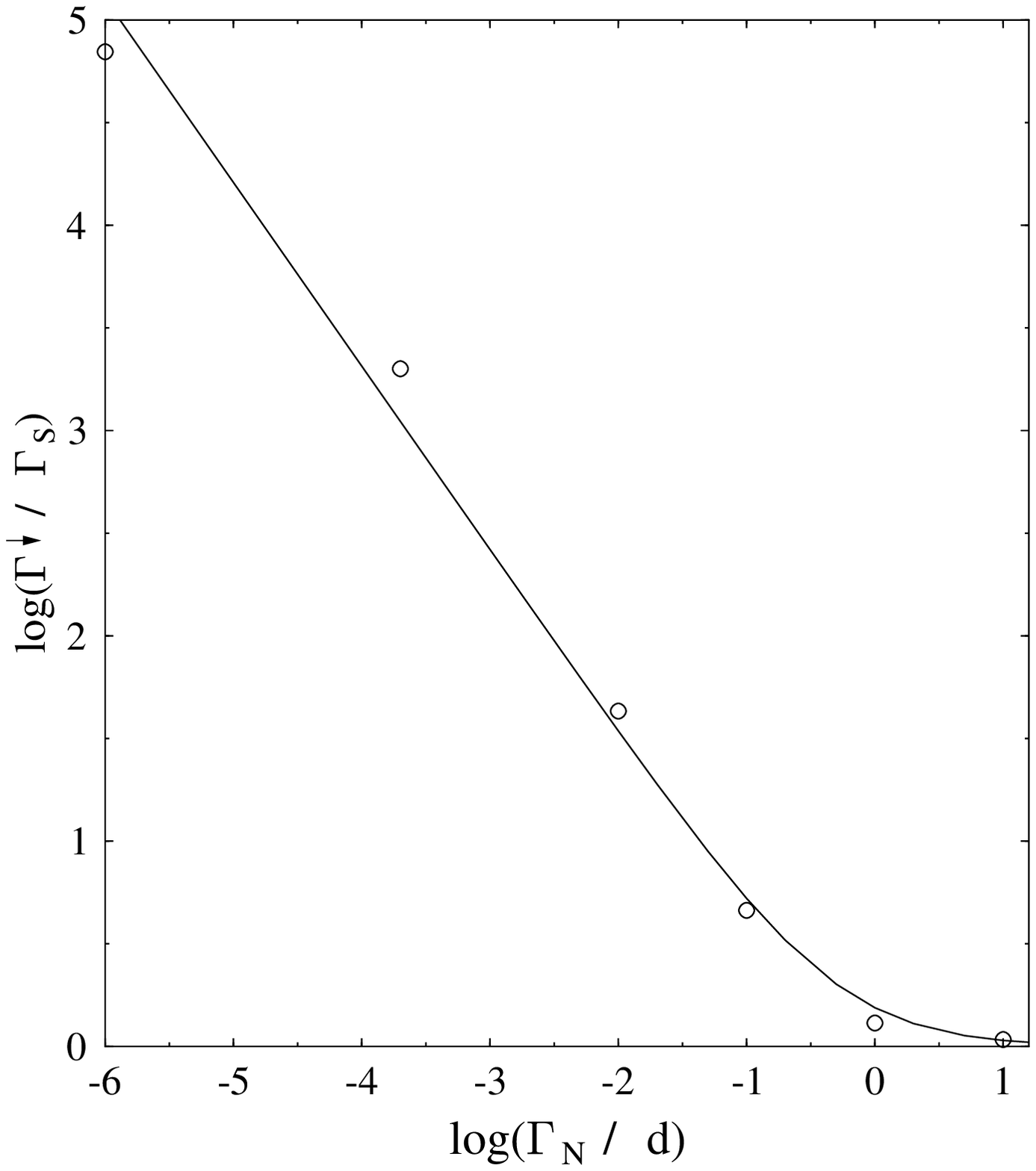,width=15cm,angle=0}}
\caption{
Comparison of the fit formula Eq.~(\ref{fit1}) (solid line) with the
calculated values of $\Gamma^{\downarrow}/\Gamma_S$ (circles) for
which the average intraband decay intensity attains the value 1/2. The basis of
logarithm is ten.}
\label{figmc}
\end{figure}

\section{Comparison with the Approach by Vigezzi et al.}
\label{vig}
 
The experimental results for decay out of a SD band have frequently been 
compared with the approach developed by Vigezzi et al. Therefore, we 
investigate in this Section how the approach and the results of 
Ref.~\cite{vig90} compare wih the theory developed in previous Sections 
of this paper.

The model which serves as starting point of the approach of Ref.~\cite{vig90} 
is identical to the one used above, see Eqs.~(\ref{ham}, \ \ref{sig}, \ 
\ref{eff}, \ \ref{in}). The formula actually used by Vigezzi et al. to 
calculate the probability $I_{\rm out}$ for decay out of the SD band is, 
however, not really derived from this model. It is rather based on physically 
plausible and intuitive reasoning which we summarize as follows. Let $| m 
\rangle$ with $m = 1, \ldots, K + 1$ denote the $(K + 1)$ eigenstates of the 
Hermitian matrix $H$ defined in Eq.~(\ref{ham}). We note that this matrix 
does not contain the decay widths $\Gamma_S$ and $\Gamma_N$ which appear in 
the width matrix, Eq.~(\ref{sig}), and in the effective Hamiltonian $H_{\rm 
eff}$, Eq.~(\ref{eff}). Let $c_m = \langle 0 | m \rangle$ be the amplitude 
with which the SD state $| 0 \rangle $ is admixed to the state $| m \rangle$. 
It is argued that the width $\Gamma_S(m)$ for decay of the state $| m \rangle$ 
into the SD band is given by $\Gamma_S(m) = |c_m|^2 \Gamma_S$. The width 
$\Gamma_N(m)$ for E1 decay of the state $| m \rangle$ is correspondingly 
written as $\Gamma_N(m) = (1 - |c_m|^2) \Gamma_N$. Finally, it is assumed that 
E2 decay out of the next--higher state in the SD band populates the state 
$| m \rangle$ with probability $|c_{m}|^{2}$. The intensity $I_{\rm out}$ for 
decay out of the SD band is then given by summing over all states $| m 
\rangle$ as
\begin{equation}
\label{out}
I_{\rm out}^{\rm vig} = \sum_m |c_m|^2 \frac{(1-|c_m|^2) \Gamma_N}{(1-
|c_m|^2) \Gamma_N + |c_m|^2 \Gamma_S} \ .
\end{equation}
We have added a superscript vig to identify the origin of this formula. We 
note that according to Eq.~(\ref{out}), the probability distribution of 
$I_{\rm out}^{\rm vig}$ and, therefore, also that of the intraband E2 
intensity $I_{\rm in}^{\rm vig} = 1 - I_{\rm out}^{\rm vig}$, depend on two 
dimensionless variables: The ratio $\Gamma_S / \Gamma_N$ which appears 
explicitly in Eq.~(\ref{out}), and the ratio $\Gamma^{\downarrow} / d$ 
which determines the statistical behavior of the mixing parameters $|c_m|^2$. 
We note that this parametric dependence of the Vigezzi model differs from 
the one characterizing the exact theory of Section~\ref{ana} where the 
relevant parameters are $\Gamma_N / d$ and $\Gamma^{\downarrow} / \Gamma_S$. 
The reasoning behind Eq.~(\ref{out}) leads us to expect that this equation 
renders a useful approximation to the exact result whenever $\Gamma_S$ and 
$\Gamma_N$ are sufficiently small (case of isolated resonances, $\Gamma_S$ 
and $\Gamma_N \ll d$). This is also suggested by the observation that 
$I_{\rm out}^{\rm vig}$ is independent of the value of the fine--structure 
constant, a result which is not physically plausible. The worrisome aspect 
is that analysis of the data using the Vigezzi approach yields values of 
$\Gamma^{\downarrow}$ which are about two orders of magnitude smaller than 
$\Gamma_N$ \cite{kru96, kru97, kue97, kho96}, thereby putting the entire 
approach into question. This is in fact what prompted the work of 
Ref.~\cite{wei98} as well as the present investigation. By comparing numerical 
results obtained for $I_{\rm in}^{\rm vig}$ with those generated from the 
model of Eqs.~(\ref{ham}, \ \ref{sig}, \ \ref{eff}, \ \ref{in}), we now 
display the limits of validity of the approach of Vigezzi et al.

Fig.~5 shows the average total intraband decay intensity calculated from
the theory of Section~\ref{ana} (solid lines) and the result of the 
approximation~(\ref{out}) (dotted lines) versus $\Gamma^{\downarrow} / 
\Gamma_S$ for six choices of $\Gamma_N / d$. We see that the difference 
between the approximation by Vigezzi et al. and the full theory becomes 
significant only at values of $\Gamma_N / d \approx 1$ which lie well 
outside the range which is relevant for the data of Refs.\cite{kru96, kru97, 
kue97, kho96} as well as the expected range of validity of the Vigezzi 
approach. The plots in Fig.~6 use the parameter $\Gamma^{\downarrow} / d$ 
of the Vigezzi model to define the abscissa and show similar behavior. Further
insight is provided by considering the case $1 \gg \Gamma^{\downarrow} / 
\Gamma_S \gg \Gamma_N / d$ which combines weak mixing between SD and ND 
states with strong fluctuations. In this case, a modified perturbation
treatment yields to lowest order
\begin{equation}
\label{vig1}
\overline{I_{\rm out}} = \sqrt{ \frac{\pi}{2} \frac{\Gamma^{\downarrow} 
\Gamma_N}{\Gamma_S d} } \ \ {\rm for} \ \ 1 \gg \Gamma^{\downarrow} / 
\Gamma_S \gg \Gamma_N / d \ .
\end{equation}
A related formula was given by Vigezzi et al. \cite{vig90}. The predictions 
of this formula are shown as dash--dotted lines in Fig.~6. We note that 
whenever the condition of validity $1 \gg \Gamma^{\downarrow} / \Gamma_S \gg 
\Gamma_N / d$ is met, the predictions of Eq.~(\ref{vig1}) agree very well 
with the exact result.

\begin{figure}
\centerline{\psfig{figure=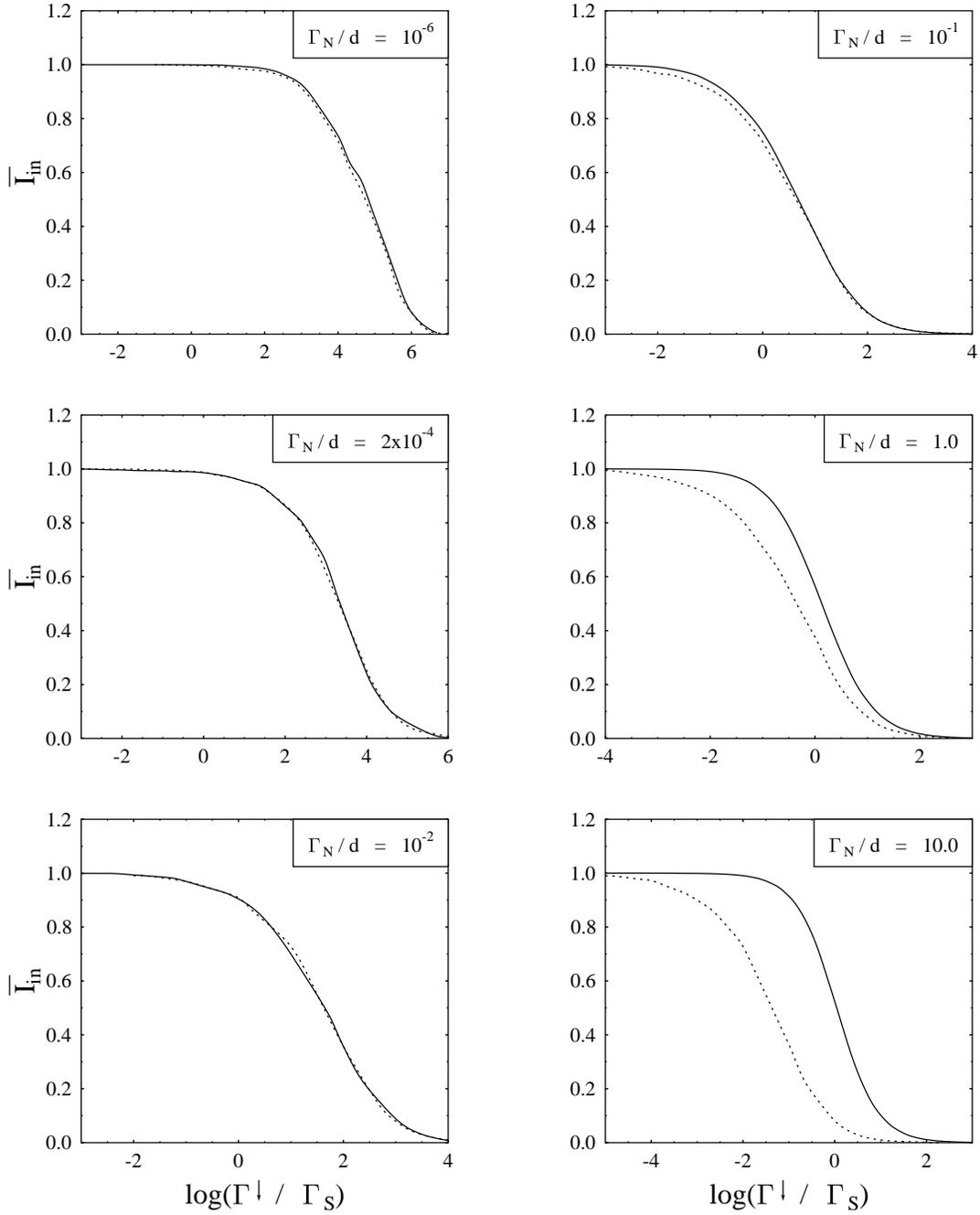,width=17cm,angle=0}}
  \caption{
Comparison of the ensemble--averaged total intraband decay intensity (solid lines) 
$\overline{I_{\rm in}}$ 
(abbreviated as $\overline{I}$)
with the prediction of the Vigezzi approach (dotted lines). The basis of logarithm
is ten.}
\label{figmd}
\end{figure}

\begin{figure}
\centerline{\psfig{figure=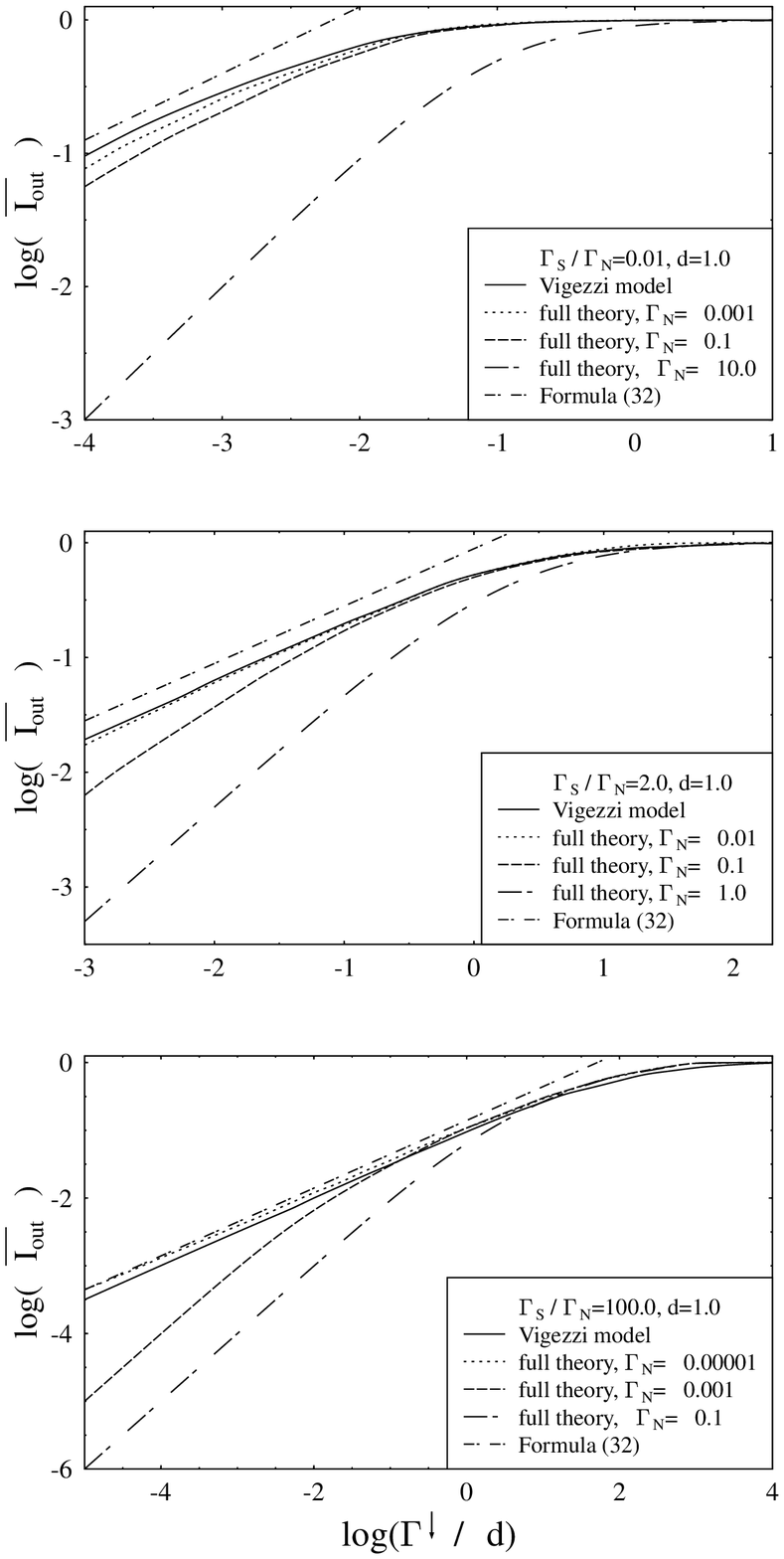,width=13cm,angle=0}}
\caption{
The logarithm of the ensemble--averaged decay intensity $\overline{I_{\rm 
out}}$ out of the SD band versus the logarithm (basis ten) of the ratio 
$\Gamma^{\downarrow} / d$ for three values of $\Gamma_S / \Gamma_N$. We show 
predictions of the Vigezzi approach (solid lines), results of the full theory 
for several values of $\Gamma_N$, and the prediction of Eq.~(\ref{vig1}). }
\label{figme}
\end{figure}

Can we determine the limits of validity of the Vigezzi approach also from 
theoretical arguments? As noted above, inspection of Eq.~(\ref{out}) suggests 
that the approach is correct to lowest non--vanishing order in both $\Gamma_S$ 
and $\Gamma_N$ (isolated resonances). Moreover, it seems to account for all 
orders of $V_{0j}$. The restriction to non--overlapping resonances becomes 
obvious by considering the case $\Gamma_N / d \geq 1$. Here, $\overline{I_{\rm 
in}}^{\rm fluc}$ is negligible, and $\overline{I_{\rm out}}$ is simply given 
by $ (\Gamma^{\downarrow} / \Gamma_S) / (1 + (\Gamma^{\downarrow} / 
\Gamma_S))$. This result is obviously and not surprisingly inaccessible to 
Eq.~(\ref{out}). For the case of strong coupling $\Gamma^{\downarrow} \gg 
\Gamma_S$, it simplifies to $\overline{I_{\rm out}} \approx 1 - (\Gamma_S / 
\Gamma^{\downarrow})$. This is in contrast to the expression $(1 - d \Gamma_S 
/ (\pi \Gamma^{\downarrow} \Gamma_N))$ derived in Ref.~\cite{vig90} for the 
same regime. Aside from this rather trivial point, a better understanding of
the domain of validity of Eq.~(\ref{out}) is obtained by considering two 
limiting cases, the limit $\Gamma^{\downarrow} \rightarrow 0$ with $\Gamma_N 
\neq 0$ fixed (limit A), and the limit $\Gamma_N \rightarrow 0$ with 
$\Gamma^{\downarrow} \neq 0$ fixed (limit B). For limit A, Eq.~(\ref{ana9}) 
shows that $\overline{I_{\rm in}^{\rm fluc}}$ vanishes as 
$(\Gamma^{\downarrow})^2$. Keeping only linear terms in $\Gamma^{\downarrow}$, 
we have from Eqs.~(\ref{ana3}) and (\ref{ana4}) that $\overline{I_{\rm in}} = 
(1 + \Gamma^{\downarrow} / \Gamma_S)^{-1} \approx 1 - \Gamma^{\downarrow} / 
\Gamma_S$ and, by Eq.~(\ref{uni}), that $\overline{I_{\rm out}} \approx 
\Gamma^{\downarrow} / \Gamma_S$. This agrees with Eq.~(\ref{ave4}). For 
limit B, we recall that for $\Gamma_N = 0$, the right--hand--side of 
Eq.~(\ref{ana9}) must equal $ (\Gamma^{\downarrow} / \Gamma_S) / (1 + 
(\Gamma^{\downarrow} / \Gamma_S))$. (Admittedly, this is not immediately 
obvious from Eq.~(\ref{ana9})). The lowest--order correction to this 
result is obtained by expanding the exponential in Eq.~(\ref{ana9}) in powers
of $\Gamma_N / d$. The first--order term is negative and yields for 
$\overline{I_{\rm out}}$ a positive term linear in $\Gamma_N / d$. While 
Eq.~(\ref{out}) is in keeping with limit B, this equation is not consistent 
with limit A: $I_{\rm out}^{\rm vig}$ never becomes independent of $\Gamma_N$, 
no matter how small $\Gamma^{\downarrow}$. This confirms the belief stated in
the Introduction that the approach of Vigezzi et al. applies only for 
sufficiently large values of $\Gamma^{\downarrow}$. To quantify this 
statement, we have calculated the terms of next order in the perturbation 
expansion of Section~\ref{pert}. We find that, for $\Gamma_N \ll d$, these are
of order $(\Gamma^{\downarrow} / \Gamma_S)^2 (d / \Gamma_N)$. Limit A is 
excluded if the second--order terms are at least of the same order as the 
first--order ones, i.e., whenever $\Gamma^{\downarrow} / \Gamma_S \geq 
\Gamma_N / d$ or $\Gamma_N \leq d (\Gamma^{\downarrow} / \Gamma_S)$. Violation 
of this condition accounts for the deviations between the exact results and 
the Vigezzi approach displayed in Fig.~6.

In conclusion, we see that the approach by Vigezzi et al. is subject to two
constraints. The obvious one is that it deals with {\it isolated} resonances. 
This implies $\Gamma_N \ll d$. The second, less obvious one is due to the 
constraint $\Gamma_N \leq d (\Gamma^{\downarrow} / \Gamma_S)$.

\section{Summary}
\label{sum}

In the present paper we have calculated the ensemble average and properties 
of the distribution function of the intraband E2 decay intensity for a 
statistical process leading to decay out of a SD band. We have shown that the 
entire distribution function depends only on the two dimensionless ratios 
$\Gamma^{\downarrow} / \Gamma_S$ and $\Gamma_N / d$. Writing the intraband 
intensity as the sum of two terms, given in terms of the average decay 
amplitude and of its fluctuating part, respectively, we have shown that 
$\overline{I}^{\rm av}$ dominates for large values of $\Gamma_N / d \geq 1$, 
while both the average fluctuating part $\overline{I}^{\rm fluc}$ and the 
fluctuations around it attain maximum values for small values of $\Gamma_N / 
d$. We have proposed two fit formulas. One permits an estimate of the average 
intraband decay intensity and the other, an estimate of the value of 
$\Gamma^{\downarrow} / \Gamma_S$ for which the average intraband decay 
intensity attains the value 1/2. We have compared our results with those of 
Vigezzi et al. We have shown that the latter approach is an approximation to 
the exact theory, and we have established its limits of validity. For 
practical purposes, the approach offers a useful approximation whenever 
$\Gamma_N / d \leq 10^{-1}$. \\

{\bf Acknowledgment} \\

We thank Thomas Rupp for providing us with the GOE generating program. One
of us (HAW) is grateful to Hanns Ludwig Harney and Ben Mottelson for 
instructive discussions. We thank A. Richter and B. R. Barrett
for a careful reading of the manuscript.

\end{document}